\documentclass[aps,prb,twocolumn,floatfix,showpacs]{revtex4}

\usepackage{amsmath}
\usepackage{graphicx}
\usepackage[normalem]{ulem}

\newcommand{\ket}[1]{\left| {#1} \right\rangle}

\newcommand{\vect}[1]{\boldsymbol{{#1}}}

\newlength{\fig}
\newlength{\smallfig}
\begin{document}

\title{Chiral spin currents and spectroscopically accessible single
  merons in quantum dots}
\author{Catherine J. Stevenson}
\author{Jordan Kyriakidis}
\email{jordan.kyriakidis@dal.ca}
\homepage{http://quantum.phys.dal.ca}
\affiliation{Department of Physics and Atmospheric Science, Dalhousie
  University, Halifax, Nova Scotia, Canada, B3H 3J5}

\begin{abstract}
  We provide unambiguous theoretical evidence for the formation of
  correlation-induced isolated merons in rotationally-symmetric
  quantum dots.  Our calculations rely on neither the
  lowest-Landau-level approximation, nor on the
  maximum-density-droplet approximation, nor on the existence of a
  spin-polarized state.  For experimentally accessible system
  parameters, unbound merons condense in the ground state at magnetic
  fields as low as $B^* = 0.2$~T and for as few as $N = 3$ confined
  fermions.  The four-fold degenerate ground-state at $B^*$
  corresponds to four orthogonal merons $\ket{QC}$ characterized by
  their topological chirality $C$ and charge $Q$.  This degeneracy is
  lifted by the Rashba and Dresselhaus spin-orbit interaction, which
  we include perturbatively, yielding spectroscopic accessibility to
  individual merons.  We further derive a closed-form expression for
  the topological chirality in the form of a chiral spin current and
  use it to both characterize our states and predict the existence of
  other topological textures in other regions of phase space, for
  example, at $N=5$.  Finally, we compare the spin textures of our
  numerically exact meron states to \textit{ansatz} wave-functions of
  merons in quantum Hall droplets and find that the ansatz
  qualitatively describes the meron states.
\end{abstract}

\date{\today}

\maketitle

\section{Introduction}
Spin states of electrons confined in semiconductor quantum dots (QDs)
are an appealing platform for quantum information science due to their
long dephasing times and
controllability~\cite{engel04:contr.spin.qubit, Taylor2005,
  Foletti2009, Laird2010}. States used for this implementation must be
long lived relative to the time-scales required to flip a qubit, and
remain coherent despite system perturbations.  Topological states are
promising candidates in that they are expected to have long
decoherence times due to their \emph{global} correlations, and be
robust against local perturbations~\cite{Das-Sarma2006}.

Skyrmions and merons are examples of topological spin textures that
are predicted to form in two-dimensional (2D) electron
systems~\cite{Sondhi1993, Moon1995Spontaneous-int}.  Skyrmion
excitations in bulk 2D systems have been predicted in the $\nu=1$
quantum Hall regime~\cite{Sondhi1993, Moon1995Spontaneous-int,
  Fertig1997Hartree-Fock-th}, and can condense into the ground state
away from $\nu=1$~\cite{Brey1995Skyrme-crystal-}.  Experimental
evidence supports the existence of skyrmion spin textures in
GaAs/AlGaAs quantum wells~\cite{Barrett1995Optically-Pumpe,
  Aifer1996Evidence-of-Sky, Khandelwal2001, Gervais2005}.

A meron can be described as half a skyrmion; it contains a central
spin oriented perpendicular to the 2D plane which transitions smoothly
into an in-plane winding away from the central spin.  Unbound merons
are not low-energy states in bulk 2D systems due to the prohibitive
exchange energy of the in-plane winding.  In finite-sized systems,
such as the QDs we are considering in the present work, the textures
are stabilized, and as we show below, they can condense in the ground
state. Recent literature predicts the formation of merons in confined
systems such as QDs in large magnetic fields~\cite{Yang2005,
  Petkovic2007Fractionalizati, Milovanovic2009Meron-ground-st}.

In this work, we use configuration-interaction techniques to study a
fully interacting QD system to provide conclusive evidence for the
formation of merons in the ground state.  We find that merons form in
systems containing as few as three particles~\cite{footnote}, at
magnetic fields as low as 0.2 T, and away from the maximum density
droplet regime.  These states are degenerate.  However, we show how
this degeneracy can be lifted, and provide a method for predicting
when merons will form in QD systems.

\section{Chirality}
\label{sec:chirality-current}
The chirality $C_n({\cal C})$ of a vector field $\vect{v}(s)$,
over a closed curve ${\cal C}$ in the direction of the unit
vector $\vect{n}$ can be defined as
\begin{equation}
  \label{eq:chirality-def}
  C_n({\cal C}) = \oint_{\cal C}  \! ds \,
  \frac{\vect{n} \cdot 
    \left(\vect{v}(s) \times \partial_s \vect{v}(s) \right)}{
    \left|
      \vect{v}_{\perp}(s)
      \right|^2},
\end{equation}
where $\vect{v}_{\perp}(s) \equiv \vect{n} \times \vect{v}(s)$ is
perpendicular to $\vect{n}$ and serves as a normalization factor; it
is the global chiral character of the vector field in which we are
chiefly interested.

We confine ourselves to two-dimensional, rotationally-symmetric QDs
centred at the origin, and we take the curve ${\cal C}$ to be a circle
about the origin with radius $r$.  Furthermore, we take our direction
$\vect{n}$ to be perpendicular to the plane of the QD.  Our vector
field is the spin density $\vect{S}(\vect{r}) \equiv \langle
\vect{\hat{S}}(r, \theta) \rangle$.  Taking the QD to lie in the
$x$-$y$ plane, Eq.~\eqref{eq:chirality-def} takes the form
\begin{equation}
  \label{eq:chirality-spin}
  C_z(r) = \frac{1}{2\pi} \int_{0}^{2 \pi} \! d\theta\, 
  \frac{
    \vect{z} \cdot \left(
      \vect{S}(r,\theta) \times \partial_\theta \vect{S}(r, \theta)
    \right)
  }{
    \left| \vect{S}_{\perp}(r, \theta)  \right|^2
  },
\end{equation}
where $\vect{S}_{\perp}(r, \theta) = \vect{z} \times \vect{S}(r,
\theta)$.

We express our spin-density operator as
\begin{equation}
  \label{eq:spin-density}
  \vect{\hat{S}}(\vect{r}) = \sum_{\sigma, \sigma'} 
  \hat{\psi}^\dagger_{\sigma}(\vect{r})
  \vect{\hat{\sigma}}_{\sigma \sigma'} \hat{\psi}_{\sigma'}(\vect{r}),
\end{equation}
where $\hat{\psi}^\dagger_{\sigma}(\vect{r})$ is the field operator
creating a fermion with spin $\sigma/2$ ($\sigma = \pm 1$) at position
$\vect{r}$ and $\vect{\hat{\sigma}}$ is the vector of Pauli matrices.
The chirality then takes the form of a chiral spin current
\begin{equation}
  \label{eq:chiral-current}
  C_z(r) =  \frac{1}{2\pi} \int_0^{2\pi} \! d\theta \, j_{\text{spin}}(r, \theta),
\end{equation}
with the chiral spin current density given by
\begin{equation}
  \label{eq:chiral-spin-density}
  j_{\text{spin}}(r, \theta) = \frac{i}{2 \left| S_+ \right|^2} \,
  \left(S_+ \partial_\theta S_+^* - S_+^* \partial_\theta S_+
  \right),
\end{equation}
where $S_+ = \langle \hat{\psi}_{\uparrow}^{\dagger}\left( \vect{r}
\right) \hat{\psi}_{\downarrow}^{\phantom{\dagger}}\left( \vect{r}
\right) \rangle$.  As an example, consider the state $|\psi\rangle =
(|L_z \, S_z\rangle + \gamma |L_z'\, S_z'\rangle) / \sqrt{2}$, where
$|L_z^{\,( \prime )} \, S_z^{\,( \prime )}\rangle$ is a general,
correlated $N$-particle state with orbital angular momentum
($z$-component) $L_z^{\,( \prime )}$ and spin angular momentum
($z$-component) $S_z^{\,( \prime )}$.  For these particular states
$|\psi\rangle$, Eq.~\eqref{eq:chiral-current} yields
\begin{equation}
  \label{eq:chirality-int-state}
  C_z(r) = (L_z'-L_z) (S_z-S_z') \delta_{|S_z - S_z'|, \, 1}.
\end{equation}
Perhaps not surprisingly, states of different orbital and spin angular
momentum must be combined to produce a global chirality.  $C_z$ is an
integer winding number whose magnitude indicates the number of full
$2\pi$ rotations along a closed curve ${\cal C}$ about the QD origin,
and whose sign indicates the sense of rotation.  In the interacting
system, such windings can spontaneously occur at points of degeneracy
in the spectrum.  (Both spin and a component of orbital angular
momentum are conserved in our system.)  We show below that this can
occur in parabolic QDs containing three strongly-interacting fermions
at a point of four-fold degeneracy.  We further show that the
spin-orbit interaction splits this degeneracy into four
spectroscopically distinct merons $\ket{QC}$ with $Q,C=\pm 1$.

\section{Theory}
\label{sec:theory}

Our system consists of $N$ interacting fermions of charge $e$, bound
to a 2D plane and laterally confined by a parabolic potential.  The 2D
Hamiltonian used to describe this ``standard model'' is
\begin{subequations}
  \label{eq:hamil}
  \begin{equation}
    \label{eq:intHam}
    \hat{\mathcal{H}}=\sum_i^N \hat{h}_i + \frac{1}{2}\sum_{i \neq j}^N
    \frac{e^2}{\epsilon | \hat{\vect{r}}_i-\hat{\vect{r}}_j |} + \hat{H}_{SO},
  \end{equation}
  where $\epsilon$ is the dielectric constant of the medium and
  $\hat{h}$ is the single-particle Hamiltonian describing harmonic
  confinement in a perpendicular magnetic field:
  \begin{equation}
    \label{eq:nonIntHam}
    \hat{h} = \frac{1}{2m^*}
    \widehat{P}^2 \! \left(\hat{\vect{r}} \right) 
    + \frac{1}{2} m^* \omega_0^2 \hat{r}^2 + g \mu_B \hat{S}_z
    B_z,
  \end{equation}
\end{subequations}
where $\hat{\vect{r}} = (\hat{x},\hat{y})$ is the position operator,
$m^*$ the effective mass, and $\omega_0$ the parabolic confinement
frequency.  We work in a finite magnetic field perpendicular to the
plane of the QD, and so $\widehat{\vect{P}} = \hat{\vect{p}} +
\frac{e}{c}\vect{A}\left(\hat{\vect{r}} \right)$, with
$\vect{A}\left(\hat{\vect{r}} \right)=\frac{B}{2}\left(-\hat{y},
  \hat{x},0 \right)$ in the symmetric gauge.  The final term in
Eq.~\eqref{eq:nonIntHam} is the Zeeman energy and plays no significant
role in what follows due to the relatively small magnitude of both the
$g$-factor ($g \approx -0.44$ in GaAs) and the magnetic field ($B <
1$~T).  The final term in Eq.~\eqref{eq:intHam} is the spin orbit
interaction, which we deal with perturbatively in a subsequent
section. In what follows, we fix the material parameters in
Eq.~\eqref{eq:hamil} to typical values for GaAs heterostructures: $m^*
= 0.067m_e$, $\epsilon = 12.4$, and $\hbar\omega_0 = 1$
meV~\cite{Hai2000Direct-determin, Kouwenhoven1991Single-electron,
  Tarucha1996Shell-filling-a}.  Finally, we take $\hbar/2=1$.

Neglecting for now the spin-orbit interaction, we obtain the
correlated eigenstates and the energies of Eq.~\eqref{eq:hamil} by
configuration-interaction methods. The single particle system,
Eq.~\eqref{eq:nonIntHam}, can be solved exactly yielding the well
known Fock-Darwin spectrum~\cite{Jacak1998:Quantum-Dots} with states
$\ket{mns}$ labelled by harmonic oscillator numbers $m,n = 0, 1, 2,
\ldots$, and spin $s = \pm 1/2$.  We use these states to build up a
basis of antisymmetrized many-body states, and write the Hamiltonian
as a block-diagonal matrix with blocks segregated by the conserved
quantities of particle number $N$, total spin $S$, $z$-component of
spin $S_z$, and orbital angular momentum $L_z$.

\begin{figure}
  \centering \resizebox{\smallfig}{!}{\includegraphics[trim = 1mm 2mm
    1mm 4mm, clip]{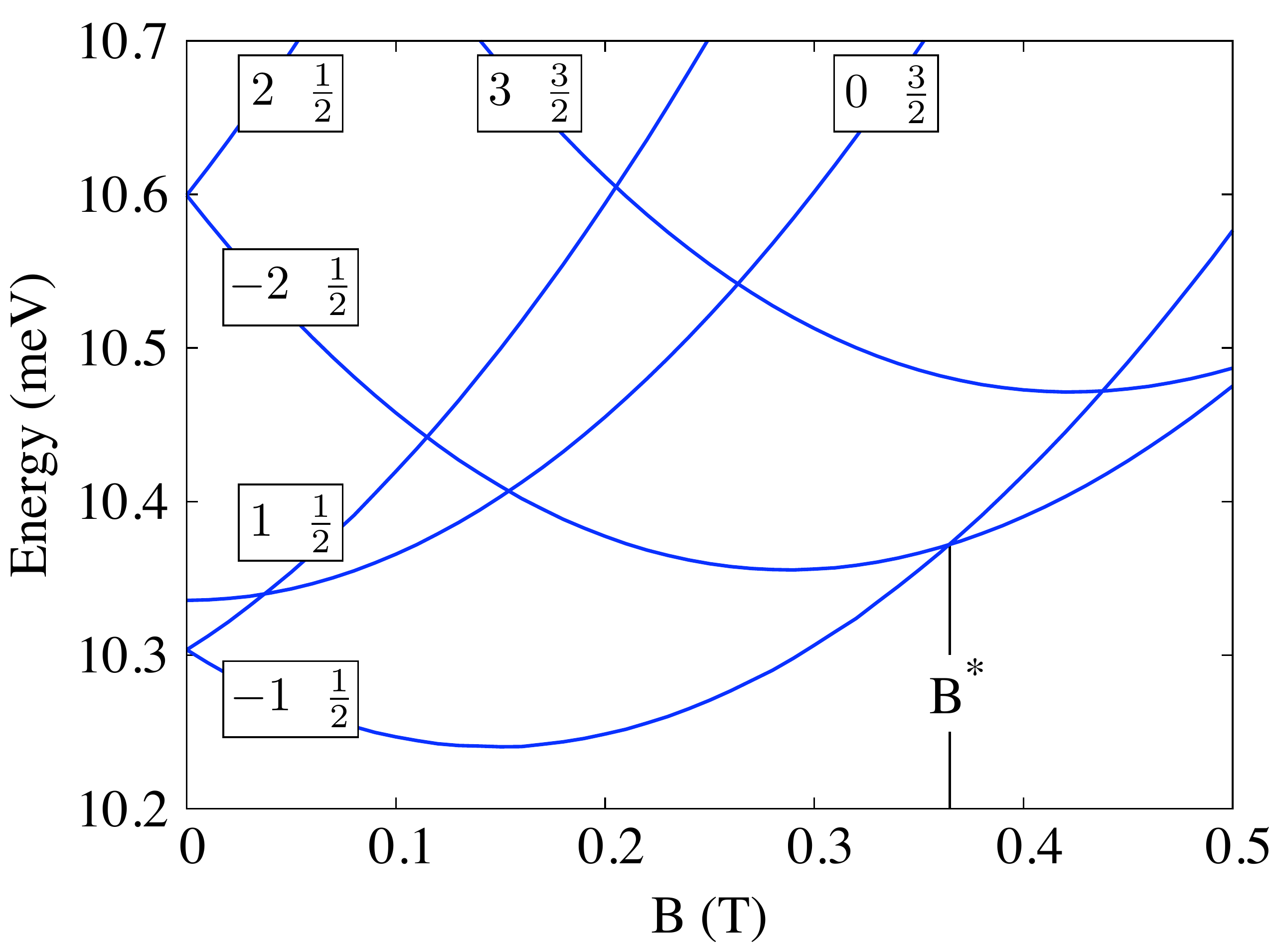}}
  \caption{\label{fig:low-lying-spectra}Low-lying spectra for $N=3$
    interacting particles with 2D harmonic confinement,
    Eq.~\eqref{eq:hamil}.  $L_z$ and $S$ quantum numbers are
    displayed for each state.  The field $B^*$ at which merons form in
    the ground state is marked.}
\end{figure}

The simplest system exhibiting meron textures occurs already for $N =
3$ confined particles.  The low-lying spectrum is shown in
Fig.~\ref{fig:low-lying-spectra}.
For the system parameters given above, a ground state degeneracy
exists at a field of $B^* = 0.365$ T.  This degenerate manifold is
spanned by the four spin-1/2 states $|L_z, S_z\rangle = |-1, \pm
1/2\rangle,\ |-2, \pm 1/2\rangle$.  In the harmonic oscillator basis,
these correlated states typically contain several thousand
Hartree-Fock states in order to reach convergence in the
energies to within 0.05\%.

\section{Meron textures}
\label{sec:meron-textures}

\begin{figure}
  \centering 
  \resizebox{\smallfig}{!}{\includegraphics{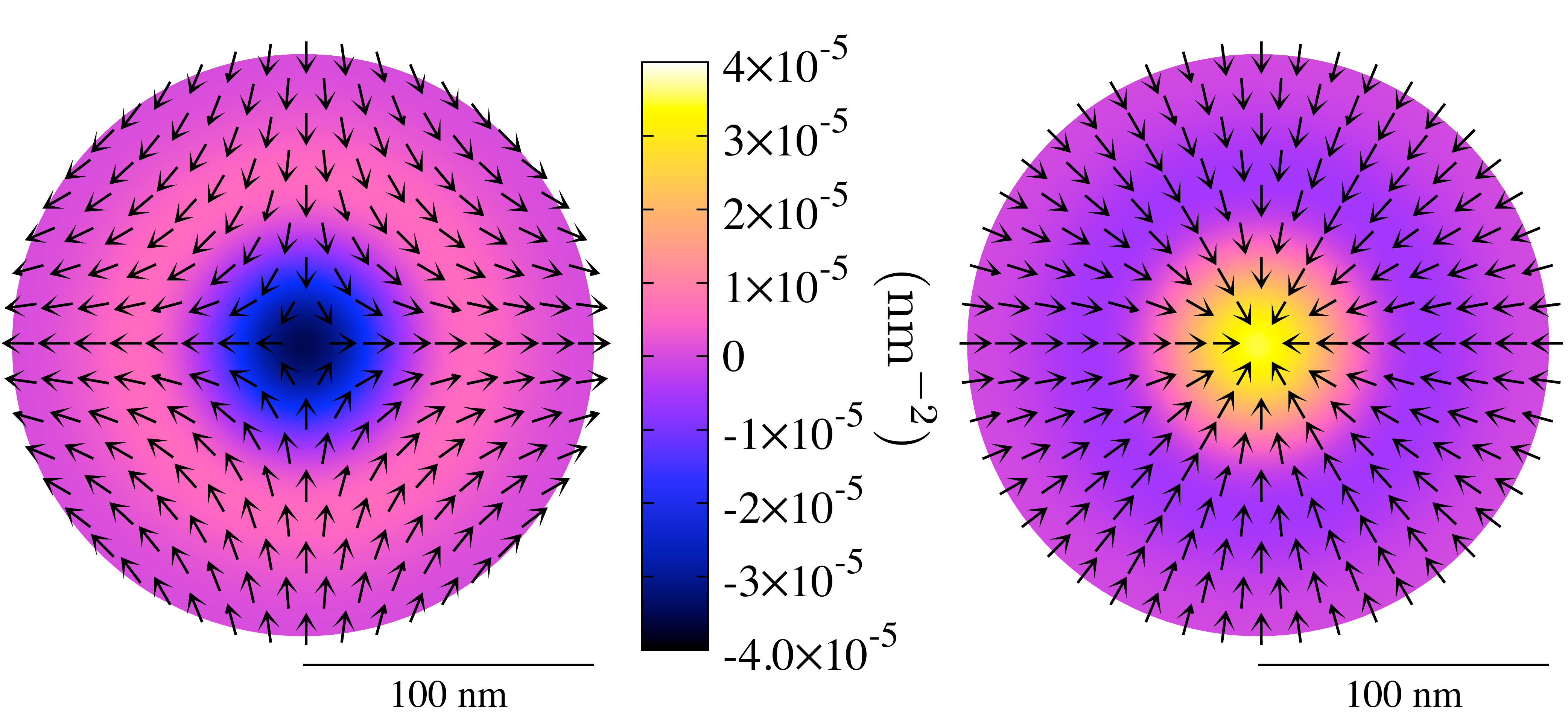}}
  \caption{\label{fig:merons}(Color online) Two of the four meron
    textures for the three-particle states given in
    Eq.~\eqref{eq:meron-states}.  The spin densities $\langle
    S_x(\vect{r})\rangle$ and $\langle S_y(\vect{r})\rangle$ are
    represented by the vector field, while $\langle
    S_z(\vect{r})\rangle$ is represented by the color bar.  The
    remaining two configurations are obtained from these by a local
    in-plane spin rotation of $\pi$.  The textures shown were
    obtained at the four-fold degeneracy point $B = B^*$ (see
    Fig.~\ref{fig:low-lying-spectra}).}
\end{figure}

The degenerate ground-state subspace at $B^*$ described above contains
merons.  An explicit form which is important in what follows is
\begin{multline}
  \label{eq:meron-states}
  |QC\rangle = 
  \frac{1 - C}{2\sqrt{2}} \left(
    \left| -1, {\textstyle \frac{1}{2}} \right\rangle 
    -i Q \left| -2, - {\textstyle \frac{1}{2}} \right\rangle
  \right)
  \\ \mbox{} +
  \frac{1 + C}{2\sqrt{2}} \left(
     \left| -1, -{\textstyle \frac{1}{2}} \right\rangle 
    + Q \left| -2, {\textstyle \frac{1}{2}} \right\rangle \right).
\end{multline}
These four states for $C = \pm 1$, $Q = \pm 1$ are orthogonal and
yield the spin textures $\langle \vect{\hat{S}} (\vect{r}) \rangle$
shown in Fig.~\ref{fig:merons}.
These spin textures are plotted using the real-space wave functions of
the 2D harmonic trap~\cite{Stevenson2011Chiral-spin-tex}.  Previous
work~\cite{Yang2005, Petkovic2007Fractionalizati,
  Milovanovic2009Meron-ground-st} has focused on lowest-Landau-level
physics or on semiclassical approximations.  The present numerically
exact results show that merons can exist far beyond the semiclassical
regime; right down to the extreme quantum limit of very few confined
particles where correlations are strongest.  In particular, the
three-particle ground states shown in Fig.~\ref{fig:merons} are
\emph{not} spin-polarized states, nor do they correspond to the $\nu =
1$ maximum density droplet.  The states in Fig.~\ref{fig:merons}
correspond to a filling factor of $2 > \nu > 1$.  (The angular
momentum of the $N$-particle maximum density droplet is $L_z =
-N(N-1)/2$.)

\begin{figure}
  \centering 
  \resizebox{\smallfig}{!}{\includegraphics{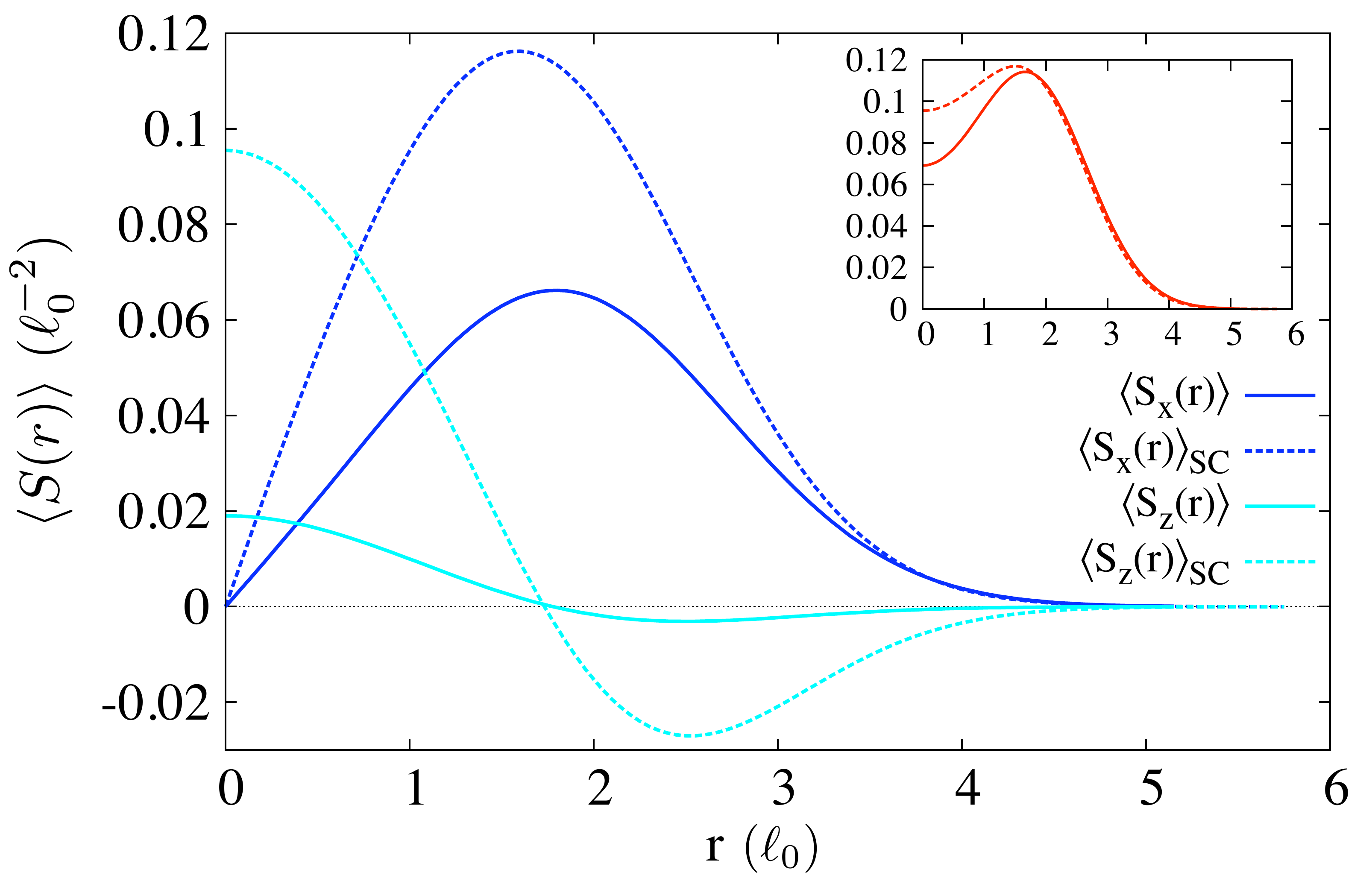}}
  \caption{\label{fig:merons-comparison}(Color online) Comparison of
    spin density components for meron $\ket{QC} = \ket{1\,1}$ of
    Eq.~\eqref{eq:meron-states} (solid lines) and the semiclassical
    result of Ref.\cite{Milovanovic2009Meron-ground-st} (dashed lines)
    for $\theta=0$.  The $S_x(r)$ curves vanish at the origin, and
    $S_y(r)$ is zero in both cases here.  Inset: Particle density for
    the $\ket{1\,1}$ meron (solid) and semiclassical (dashed) result.}
\end{figure}

A semiclassical (SC) \textit{ansatz} for a single meron in a quantum
Hall droplet is given in Ref.~\cite{Milovanovic2009Meron-ground-st} in
terms of spinors and Laughlin wave functions.
Figure~\ref{fig:merons-comparison} shows a direct comparison between
the spin textures $\langle \vect{S}(\vect{r})\rangle$ for the
$\ket{QC}=\ket{1\,1}$ meron presented here and the SC \textit{ansatz}
for $N=3$.  Results are shown in units of $\ell_0 \equiv
\sqrt{\hbar/2m^*\omega}$, where $\omega \equiv
\sqrt{\omega_0^2+\omega_c^2/4}$, and $\omega_c$ is the cyclotron
frequency.  The two chief distinctions between the curves are the
relative suppression of both charge (primarily at the origin) and spin
polarization throughout the dot in the present work.  Both effects can
be attributed to the strong correlation effects captured in the
present treatment.  The suppression of charge at the origin is a
result of the greater repulsion among the particles.  The
correlation-induced mixing among numerous Landau-levels and
angular-momentum and spin-resolved Slater determinants likewise has
the effect of reducing the spin polarization throughout the dot.  Indeed
we find a conspicuous lack of full polarization throughout the dot.

\section{Spin-Orbit Coupling}
\label{sec:SO}
To linear order, the spin-orbit interaction splits the degeneracy
among the four merons states $\ket{QC}$.  Remarkably, we find the
Rashba term couples only to merons with positive chirality, whereas
the Dresselhaus term couples only to those with negative chirality.
Explicitly, the linearized spin-orbit interaction may be written as
~\cite{DRESSELHAUS1955SPIN-ORBIT-COUP, BYCHKOV1984PROPERTIES-OF-A,
  Dyakonov1972Spin-relaxation, SILVA1994SPIN-SPLIT-SUBB,
  Silva1997Spin-orbit-spli, Sousa2003Gate-control-of,
  Florescu2006Spin-relaxation}
\begin{equation}
  \label{eq:HSO}
  \hat{H}_{SO} = \beta \left( -\sigma_x \text{P}_x + \sigma_y \text{P}_y\right)
  + \alpha \left( \sigma_x \text{P}_y - \sigma_y \text{P}_x \right),
\end{equation}
where $\alpha$ and $\beta$ are the Rashba and Dresselhaus spin-orbit
coupling strengths, respectively.  Within the lowest-energy degenerate
subspace, the spin-orbit-induced energy splittings E$_{QC}$ associated
with each state $|QC\rangle$ are,
\begin{equation}\label{eq:SOenergies}
  \text{E}_{QC} = \frac{{\cal{E}}_C Q}{2}\left[ (1+C)\alpha + 
    (1-C)\beta \right],
\end{equation}
where,
\begin{equation}
  {\cal{E}}_C = \lambda \langle -2, \, \tfrac{C}{2}|
  \left[ (1-\tfrac{1}{\nu}) \hat{S}_C\,\hat{b}^\dagger - (1+\tfrac{1}{\nu}) 
    \hat{S}_C\,\hat{a}  \right] |-1, \, -\tfrac{C}{2} \rangle,
\end{equation}
with $\lambda=1/(\sqrt{2} \ell_0)$ and $Q,C=\pm$.  Here, $\hat{a}$ and
$\hat{b}^\dagger$ independently lower the orbital angular momentum of
the $|L_z, S_z\rangle$ states, while $\hat{S}_{\pm}$ raises or lowers
the spin.  Also, $\nu=\sqrt{1+4\omega_0^2 / \omega_c^2}$.  Note if
either $\alpha$ or $\beta$ is zero, or if $\alpha=\beta$, a two-fold
degeneracy remains.  Otherwise, the degeneracy is completely lifted.
Since the Rashba term is to some extent tunable through
externally-applied gate voltages~\cite{Nitta1997Gate-control-of,
  Engels1997Experimental-an, Studer2009Gate-Controlled},
Eq.~\eqref{eq:SOenergies} demonstrates a measure of experimental
control over the merons.

\section{Predictions}
\label{sec:predict}
\begin{figure}
  \centering 
  \resizebox{\smallfig}{!}{\includegraphics{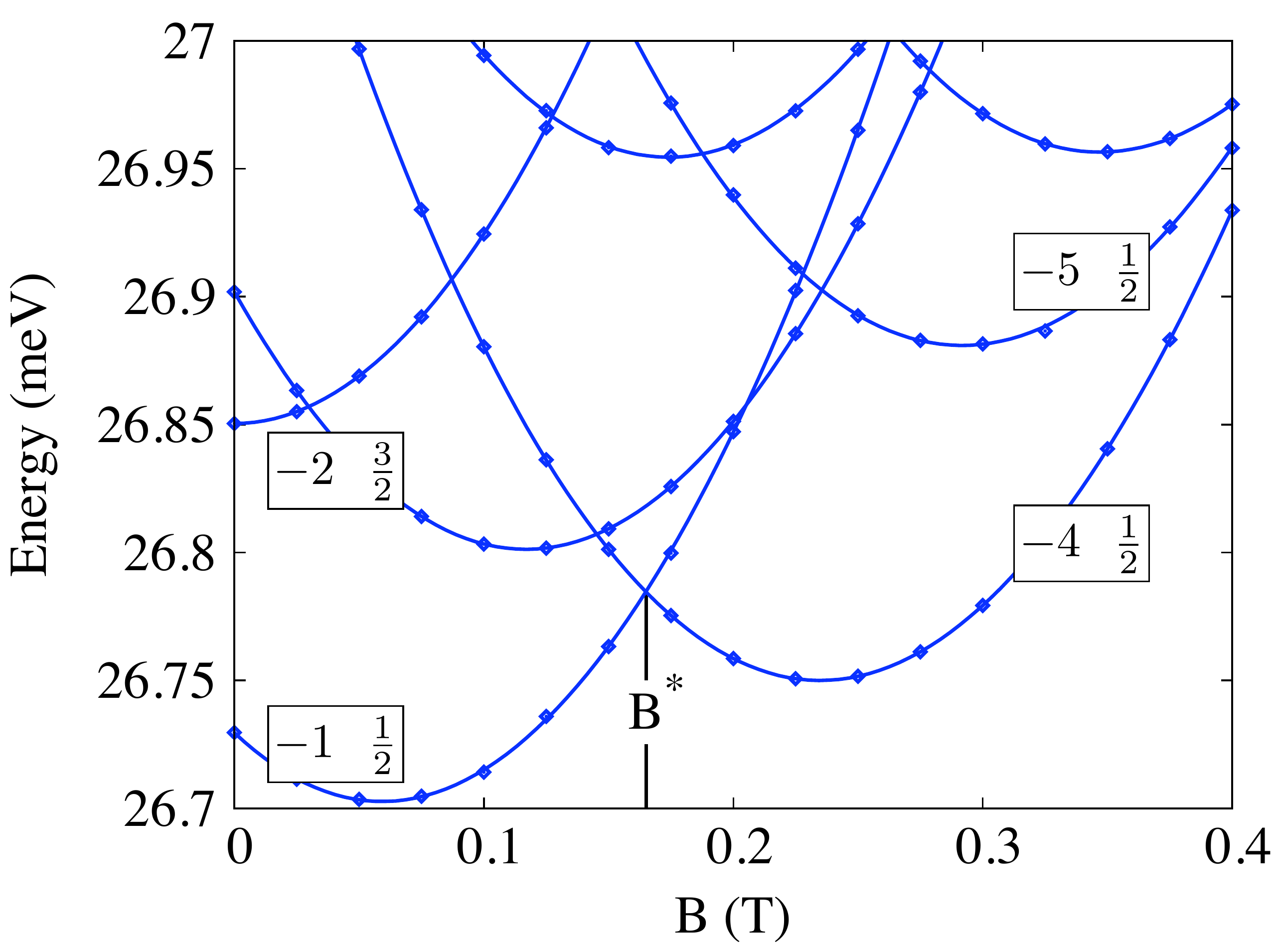}}
  \caption{\label{fig:N5-spectra}(Color online) Low-lying spectra for
    $N=5$ interacting particles with 2D harmonic confinement,
    Eq.~\eqref{eq:hamil}.  The quantum numbers $L_z$ and $S$ of
    selected levels are shown.  $B^*$ marks the field at which merons form
    in the ground state.}
\end{figure}

\begin{figure}
  \centering 
  \resizebox{\smallfig}{!}{\includegraphics{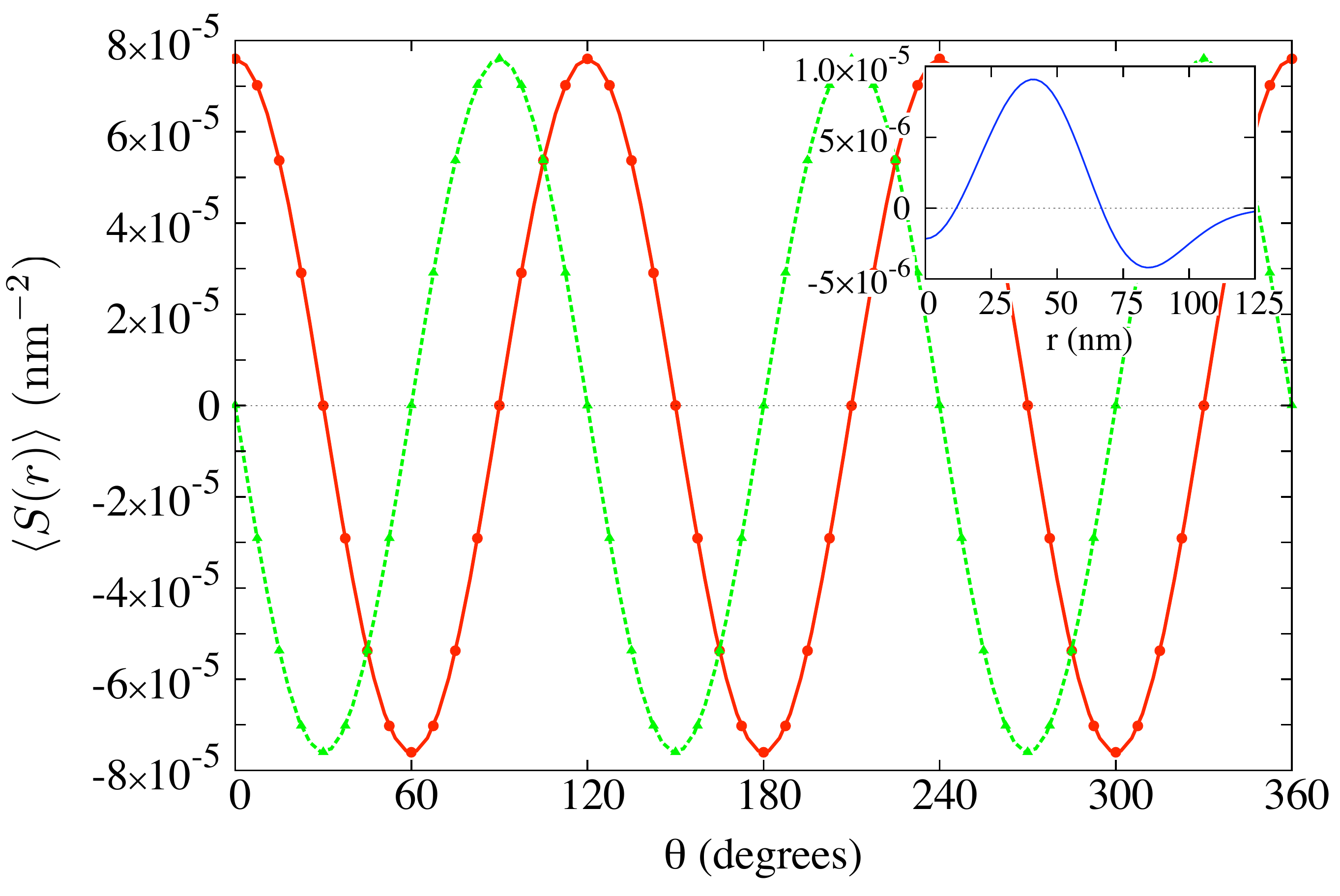}}
  \caption{\label{fig:N5-winding}(Color online) Meron spin texture
    $\ket{QC} = \ket{1\,-3}$ in the five-particle system at the
    degeneracy point $B^*$ (Fig.~\ref{fig:N5-spectra}).  Plotted is
    $\langle S_x(r, \theta)\rangle$ (solid) and $\langle
    S_y(r,\theta)\rangle$ (dashed) around a closed curve at radius
    $r=67$ nm.  The winding number of 3 does not depend on $r$.
    Rather, the amplitudes decay with $r$ due to the finite extent of
    the system.  Inset: $\langle S_z(r)\rangle$ dependence on radius.}
\end{figure}

Equation~\eqref{eq:chirality-int-state} predicts that degenerate
manifolds with states of different orbital and spin angular momentum
contain quasi-topological winding spin textures. Furthermore, the
magnitude of the winding number associated with these states is
equal to the difference between the orbital angular momentum of the
degenerate states.  In the case of a four-fold degeneracy,
Eq.~\eqref{eq:meron-states} generalizes to
\begin{multline}
  \label{eq:general-meron}
  |QC\rangle = 
  \frac{1 - c}{2\sqrt{2}} \left(
    \left| L_z, S_z \right\rangle 
    -i q \left| L_z', S_z' \right\rangle
  \right)
  \\ \mbox{} +
  \frac{1 + c}{2\sqrt{2}} \left(
     \left| L_z, S_z' \right\rangle 
    + q \left| L_z', S_z \right\rangle \right),
\end{multline}
where $L_z \neq L_z'$, $|S_z - S_z'|= 1$, $c \equiv C / |C|$, and $q
\equiv Q / |Q|$.  The existence criteria for the quasi-topological
winding textures described by Eq.~\eqref{eq:general-meron} is
satisfied throughout the phase space of the QD.  For example,
Fig.~\ref{fig:N5-spectra} displays the lowest-lying spectrum for the
five-particle system using the same experimental parameters listed
above.
The four-fold degeneracy at $B^* = 0.16$ T occurs between the states
$|L_z, S, S_z \rangle = |-1, 1/2, \pm 1/2\rangle$ and $|-4, 1/2, \pm
1/2\rangle$.  According to Eq.~\eqref{eq:chirality-int-state}, the
quasi-topological winding states which exist at this point have a
winding of three.  Indeed this can be seen in
Fig.~\ref{fig:N5-winding}, where the individual spin components for a
meron state with $QC$ quantum numbers $\ket{QC}=\ket{1\,-3}$ are
plotted.
The large, off-centre peak in the $\langle S_z(r)\rangle$ distribution
is due to the increased Coulomb interaction strength relative to the
three-particle system.

\begin{acknowledgments}
This work was supported by the Natural Science and Engineering
Research Council of Canada, by the Canadian Foundation for Innovation,
and by the Lockheed Martin Corporation.
\end{acknowledgments}

\bibliography{Stevenson2011}

\begin{thebibliography}{32}
\expandafter\ifx\csname natexlab\endcsname\relax\def\natexlab#1{#1}\fi
\expandafter\ifx\csname bibnamefont\endcsname\relax
  \def\bibnamefont#1{#1}\fi
\expandafter\ifx\csname bibfnamefont\endcsname\relax
  \def\bibfnamefont#1{#1}\fi
\expandafter\ifx\csname citenamefont\endcsname\relax
  \def\citenamefont#1{#1}\fi
\expandafter\ifx\csname url\endcsname\relax
  \def\url#1{\texttt{#1}}\fi
\expandafter\ifx\csname urlprefix\endcsname\relax\def\urlprefix{URL }\fi
\providecommand{\bibinfo}[2]{#2}
\providecommand{\eprint}[2][]{\url{#2}}

\bibitem[{\citenamefont{Engel et~al.}(2004)\citenamefont{Engel, Kouwenhoven,
  Loss, and Marcus}}]{engel04:contr.spin.qubit}
\bibinfo{author}{\bibfnamefont{H.~A.} \bibnamefont{Engel}},
  \bibinfo{author}{\bibfnamefont{L.~P.} \bibnamefont{Kouwenhoven}},
  \bibinfo{author}{\bibfnamefont{D.}~\bibnamefont{Loss}}, \bibnamefont{and}
  \bibinfo{author}{\bibfnamefont{C.~M.} \bibnamefont{Marcus}},
  \bibinfo{journal}{Quantum Inf. Process.} \textbf{\bibinfo{volume}{3}},
  \bibinfo{pages}{115} (\bibinfo{year}{2004}).

\bibitem[{\citenamefont{Taylor et~al.}(2005)\citenamefont{Taylor, Engel, Dur,
  Yacoby, Marcus, Zoller, and Lukin}}]{Taylor2005}
\bibinfo{author}{\bibfnamefont{J.~M.} \bibnamefont{Taylor}},
  \bibinfo{author}{\bibfnamefont{H.~A.} \bibnamefont{Engel}},
  \bibinfo{author}{\bibfnamefont{W.}~\bibnamefont{Dur}},
  \bibinfo{author}{\bibfnamefont{A.}~\bibnamefont{Yacoby}},
  \bibinfo{author}{\bibfnamefont{C.~M.} \bibnamefont{Marcus}},
  \bibinfo{author}{\bibfnamefont{P.}~\bibnamefont{Zoller}}, \bibnamefont{and}
  \bibinfo{author}{\bibfnamefont{M.~D.} \bibnamefont{Lukin}},
  \bibinfo{journal}{Nature Phys.} \textbf{\bibinfo{volume}{1}},
  \bibinfo{pages}{177} (\bibinfo{year}{2005}).

\bibitem[{\citenamefont{Foletti et~al.}(2009)\citenamefont{Foletti, Bluhm,
  Mahalu, Umansky, and Yacoby}}]{Foletti2009}
\bibinfo{author}{\bibfnamefont{S.}~\bibnamefont{Foletti}},
  \bibinfo{author}{\bibfnamefont{H.}~\bibnamefont{Bluhm}},
  \bibinfo{author}{\bibfnamefont{D.}~\bibnamefont{Mahalu}},
  \bibinfo{author}{\bibfnamefont{V.}~\bibnamefont{Umansky}}, \bibnamefont{and}
  \bibinfo{author}{\bibfnamefont{A.}~\bibnamefont{Yacoby}},
  \bibinfo{journal}{Nature Phys.} \textbf{\bibinfo{volume}{5}},
  \bibinfo{pages}{903} (\bibinfo{year}{2009}).

\bibitem[{\citenamefont{Laird et~al.}(2010)\citenamefont{Laird, Taylor,
  DiVincenzo, Marcus, Hanson, and Gossard}}]{Laird2010}
\bibinfo{author}{\bibfnamefont{E.~A.} \bibnamefont{Laird}},
  \bibinfo{author}{\bibfnamefont{J.~M.} \bibnamefont{Taylor}},
  \bibinfo{author}{\bibfnamefont{D.~P.} \bibnamefont{DiVincenzo}},
  \bibinfo{author}{\bibfnamefont{C.~M.} \bibnamefont{Marcus}},
  \bibinfo{author}{\bibfnamefont{M.~P.} \bibnamefont{Hanson}},
  \bibnamefont{and} \bibinfo{author}{\bibfnamefont{A.~C.}
  \bibnamefont{Gossard}}, \bibinfo{journal}{Phys. Rev. B}
  \textbf{\bibinfo{volume}{82}}, \bibinfo{pages}{075403}
  (\bibinfo{year}{2010}).

\bibitem[{\citenamefont{Das~Sarma et~al.}(2006)\citenamefont{Das~Sarma,
  Freedman, and Nayak}}]{Das-Sarma2006}
\bibinfo{author}{\bibfnamefont{S.}~\bibnamefont{Das~Sarma}},
  \bibinfo{author}{\bibfnamefont{M.}~\bibnamefont{Freedman}}, \bibnamefont{and}
  \bibinfo{author}{\bibfnamefont{C.}~\bibnamefont{Nayak}},
  \bibinfo{journal}{Phys. Today} \textbf{\bibinfo{volume}{59}},
  \bibinfo{pages}{32} (\bibinfo{year}{2006}).

\bibitem[{\citenamefont{Sondhi et~al.}(1993)\citenamefont{Sondhi, Karlhede,
  Kivelson, and Rezayi}}]{Sondhi1993}
\bibinfo{author}{\bibfnamefont{S.~L.} \bibnamefont{Sondhi}},
  \bibinfo{author}{\bibfnamefont{A.}~\bibnamefont{Karlhede}},
  \bibinfo{author}{\bibfnamefont{S.~A.} \bibnamefont{Kivelson}},
  \bibnamefont{and} \bibinfo{author}{\bibfnamefont{E.~H.}
  \bibnamefont{Rezayi}}, \bibinfo{journal}{Phys. Rev. B}
  \textbf{\bibinfo{volume}{47}}, \bibinfo{pages}{16419} (\bibinfo{year}{1993}).

\bibitem[{\citenamefont{Moon et~al.}(1995)\citenamefont{Moon, Mori, Yang,
  Girvin, MacDonald, Zheng, Yoshioka, and Zhang}}]{Moon1995Spontaneous-int}
\bibinfo{author}{\bibfnamefont{K.}~\bibnamefont{Moon}},
  \bibinfo{author}{\bibfnamefont{H.}~\bibnamefont{Mori}},
  \bibinfo{author}{\bibfnamefont{K.}~\bibnamefont{Yang}},
  \bibinfo{author}{\bibfnamefont{S.~M.} \bibnamefont{Girvin}},
  \bibinfo{author}{\bibfnamefont{A.~H.} \bibnamefont{MacDonald}},
  \bibinfo{author}{\bibfnamefont{L.}~\bibnamefont{Zheng}},
  \bibinfo{author}{\bibfnamefont{D.}~\bibnamefont{Yoshioka}}, \bibnamefont{and}
  \bibinfo{author}{\bibfnamefont{S.~C.} \bibnamefont{Zhang}},
  \bibinfo{journal}{Phys. Rev. B} \textbf{\bibinfo{volume}{51}},
  \bibinfo{pages}{5138} (\bibinfo{year}{1995}).

\bibitem[{\citenamefont{Fertig et~al.}(1997)\citenamefont{Fertig, Brey, Cote,
  MacDonald, Karlhede, and Sondhi}}]{Fertig1997Hartree-Fock-th}
\bibinfo{author}{\bibfnamefont{H.~A.} \bibnamefont{Fertig}},
  \bibinfo{author}{\bibfnamefont{L.}~\bibnamefont{Brey}},
  \bibinfo{author}{\bibfnamefont{R.}~\bibnamefont{Cote}},
  \bibinfo{author}{\bibfnamefont{A.~H.} \bibnamefont{MacDonald}},
  \bibinfo{author}{\bibfnamefont{A.}~\bibnamefont{Karlhede}}, \bibnamefont{and}
  \bibinfo{author}{\bibfnamefont{S.~L.} \bibnamefont{Sondhi}},
  \bibinfo{journal}{Phys. Rev. B} \textbf{\bibinfo{volume}{55}},
  \bibinfo{pages}{10671} (\bibinfo{year}{1997}).

\bibitem[{\citenamefont{Brey et~al.}(1995)\citenamefont{Brey, Fertig,
  C{\^o}t{\'e}, and MacDonald}}]{Brey1995Skyrme-crystal-}
\bibinfo{author}{\bibfnamefont{L.}~\bibnamefont{Brey}},
  \bibinfo{author}{\bibfnamefont{H.~A.} \bibnamefont{Fertig}},
  \bibinfo{author}{\bibfnamefont{R.}~\bibnamefont{C{\^o}t{\'e}}},
  \bibnamefont{and} \bibinfo{author}{\bibfnamefont{A.~H.}
  \bibnamefont{MacDonald}}, \bibinfo{journal}{Phys. Rev. Lett.}
  \textbf{\bibinfo{volume}{75}}, \bibinfo{pages}{2562} (\bibinfo{year}{1995}).

\bibitem[{\citenamefont{Barrett et~al.}(1995)\citenamefont{Barrett, Dabbagh,
  Pfeiffer, West, and Tycko}}]{Barrett1995Optically-Pumpe}
\bibinfo{author}{\bibfnamefont{S.~E.} \bibnamefont{Barrett}},
  \bibinfo{author}{\bibfnamefont{G.}~\bibnamefont{Dabbagh}},
  \bibinfo{author}{\bibfnamefont{L.~N.} \bibnamefont{Pfeiffer}},
  \bibinfo{author}{\bibfnamefont{K.~W.} \bibnamefont{West}}, \bibnamefont{and}
  \bibinfo{author}{\bibfnamefont{R.}~\bibnamefont{Tycko}},
  \bibinfo{journal}{Phys. Rev. Lett.} \textbf{\bibinfo{volume}{74}},
  \bibinfo{pages}{5112} (\bibinfo{year}{1995}).

\bibitem[{\citenamefont{Aifer et~al.}(1996)\citenamefont{Aifer, Goldberg, and
  Broido}}]{Aifer1996Evidence-of-Sky}
\bibinfo{author}{\bibfnamefont{E.~H.} \bibnamefont{Aifer}},
  \bibinfo{author}{\bibfnamefont{B.~B.} \bibnamefont{Goldberg}},
  \bibnamefont{and} \bibinfo{author}{\bibfnamefont{D.~A.}
  \bibnamefont{Broido}}, \bibinfo{journal}{Phys. Rev. Lett.}
  \textbf{\bibinfo{volume}{76}}, \bibinfo{pages}{680} (\bibinfo{year}{1996}).

\bibitem[{\citenamefont{Khandelwal et~al.}(2001)\citenamefont{Khandelwal,
  Dementyev, Kuzma, Barrett, Pfeiffer, and West}}]{Khandelwal2001}
\bibinfo{author}{\bibfnamefont{P.}~\bibnamefont{Khandelwal}},
  \bibinfo{author}{\bibfnamefont{A.~E.} \bibnamefont{Dementyev}},
  \bibinfo{author}{\bibfnamefont{N.~N.} \bibnamefont{Kuzma}},
  \bibinfo{author}{\bibfnamefont{S.~E.} \bibnamefont{Barrett}},
  \bibinfo{author}{\bibfnamefont{L.~N.} \bibnamefont{Pfeiffer}},
  \bibnamefont{and} \bibinfo{author}{\bibfnamefont{K.~W.} \bibnamefont{West}},
  \bibinfo{journal}{Phys. Rev. Lett.} \textbf{\bibinfo{volume}{86}},
  \bibinfo{pages}{5353} (\bibinfo{year}{2001}).

\bibitem[{\citenamefont{Gervais et~al.}(2005)\citenamefont{Gervais, Stormer,
  Tsui, Kuhns, Moulton, Reyes, Pfeiffer, Baldwin, and West}}]{Gervais2005}
\bibinfo{author}{\bibfnamefont{G.}~\bibnamefont{Gervais}},
  \bibinfo{author}{\bibfnamefont{H.~L.} \bibnamefont{Stormer}},
  \bibinfo{author}{\bibfnamefont{D.~C.} \bibnamefont{Tsui}},
  \bibinfo{author}{\bibfnamefont{P.~L.} \bibnamefont{Kuhns}},
  \bibinfo{author}{\bibfnamefont{W.~G.} \bibnamefont{Moulton}},
  \bibinfo{author}{\bibfnamefont{A.~P.} \bibnamefont{Reyes}},
  \bibinfo{author}{\bibfnamefont{L.~N.} \bibnamefont{Pfeiffer}},
  \bibinfo{author}{\bibfnamefont{K.~W.} \bibnamefont{Baldwin}},
  \bibnamefont{and} \bibinfo{author}{\bibfnamefont{K.~W.} \bibnamefont{West}},
  \bibinfo{journal}{Phys. Rev. Lett.} \textbf{\bibinfo{volume}{94}},
  \bibinfo{pages}{196803} (\bibinfo{year}{2005}).

\bibitem[{\citenamefont{Yang et~al.}(2005)\citenamefont{Yang, Hwang, and
  Park}}]{Yang2005}
\bibinfo{author}{\bibfnamefont{S.-R.~E.} \bibnamefont{Yang}},
  \bibinfo{author}{\bibfnamefont{N.~Y.} \bibnamefont{Hwang}}, \bibnamefont{and}
  \bibinfo{author}{\bibfnamefont{S.}~\bibnamefont{Park}},
  \bibinfo{journal}{Phys. Rev. B} \textbf{\bibinfo{volume}{72}},
  \bibinfo{pages}{165337} (\bibinfo{year}{2005}).

\bibitem[{\citenamefont{Petkovic and
  Milovanovic}(2007)}]{Petkovic2007Fractionalizati}
\bibinfo{author}{\bibfnamefont{A.}~\bibnamefont{Petkovic}} \bibnamefont{and}
  \bibinfo{author}{\bibfnamefont{M.~V.} \bibnamefont{Milovanovic}},
  \bibinfo{journal}{Phys. Rev. Lett.} \textbf{\bibinfo{volume}{98}},
  \bibinfo{pages}{066808} (\bibinfo{year}{2007}).

\bibitem[{\citenamefont{Milovanovic et~al.}(2009)\citenamefont{Milovanovic,
  Dobardzic, and Radovic}}]{Milovanovic2009Meron-ground-st}
\bibinfo{author}{\bibfnamefont{M.~V.} \bibnamefont{Milovanovic}},
  \bibinfo{author}{\bibfnamefont{E.}~\bibnamefont{Dobardzic}},
  \bibnamefont{and} \bibinfo{author}{\bibfnamefont{Z.}~\bibnamefont{Radovic}},
  \bibinfo{journal}{Phys. Rev. B} \textbf{\bibinfo{volume}{80}},
  \bibinfo{pages}{125305} (\bibinfo{year}{2009}).

\bibitem[{foo()}]{footnote}
\bibinfo{note}{For finely tuned spin-orbit parameters, merons can form at the
  four-fold degenerate singlet-triplet point in the $N=2$ system, however only
  two of the four states have windings.}

\bibitem[{\citenamefont{Hai et~al.}(2000)\citenamefont{Hai, Chen, Buyanova,
  Xin, and Tu}}]{Hai2000Direct-determin}
\bibinfo{author}{\bibfnamefont{P.~N.} \bibnamefont{Hai}},
  \bibinfo{author}{\bibfnamefont{W.~M.} \bibnamefont{Chen}},
  \bibinfo{author}{\bibfnamefont{I.~A.} \bibnamefont{Buyanova}},
  \bibinfo{author}{\bibfnamefont{H.~P.} \bibnamefont{Xin}}, \bibnamefont{and}
  \bibinfo{author}{\bibfnamefont{C.~W.} \bibnamefont{Tu}},
  \bibinfo{journal}{Appl. Phys. Lett.} \textbf{\bibinfo{volume}{77}},
  \bibinfo{pages}{1843} (\bibinfo{year}{2000}).

\bibitem[{\citenamefont{Kouwenhoven et~al.}(1991)\citenamefont{Kouwenhoven,
  Vandervaart, Johnson, Kool, Harmans, Williamson, Staring, and
  Foxon}}]{Kouwenhoven1991Single-electron}
\bibinfo{author}{\bibfnamefont{L.~P.} \bibnamefont{Kouwenhoven}},
  \bibinfo{author}{\bibfnamefont{N.~C.} \bibnamefont{Vandervaart}},
  \bibinfo{author}{\bibfnamefont{A.~T.} \bibnamefont{Johnson}},
  \bibinfo{author}{\bibfnamefont{W.}~\bibnamefont{Kool}},
  \bibinfo{author}{\bibfnamefont{C.~J. P.~M.} \bibnamefont{Harmans}},
  \bibinfo{author}{\bibfnamefont{J.~G.} \bibnamefont{Williamson}},
  \bibinfo{author}{\bibfnamefont{A.~A.~M.} \bibnamefont{Staring}},
  \bibnamefont{and} \bibinfo{author}{\bibfnamefont{C.~T.} \bibnamefont{Foxon}},
  \bibinfo{journal}{Z. Phys. B} \textbf{\bibinfo{volume}{85}},
  \bibinfo{pages}{367} (\bibinfo{year}{1991}).

\bibitem[{\citenamefont{Tarucha et~al.}(1996)\citenamefont{Tarucha, Austing,
  Honda, van~der Hage, and Kouwenhoven}}]{Tarucha1996Shell-filling-a}
\bibinfo{author}{\bibfnamefont{S.}~\bibnamefont{Tarucha}},
  \bibinfo{author}{\bibfnamefont{D.~G.} \bibnamefont{Austing}},
  \bibinfo{author}{\bibfnamefont{T.}~\bibnamefont{Honda}},
  \bibinfo{author}{\bibfnamefont{R.~J.} \bibnamefont{van~der Hage}},
  \bibnamefont{and} \bibinfo{author}{\bibfnamefont{L.~P.}
  \bibnamefont{Kouwenhoven}}, \bibinfo{journal}{Phys. Rev. Lett.}
  \textbf{\bibinfo{volume}{77}}, \bibinfo{pages}{3613} (\bibinfo{year}{1996}).

\bibitem[{\citenamefont{Jacak et~al.}(1998)\citenamefont{Jacak, Hawrylak, and
  Wojs}}]{Jacak1998:Quantum-Dots}
\bibinfo{author}{\bibfnamefont{L.}~\bibnamefont{Jacak}},
  \bibinfo{author}{\bibfnamefont{P.}~\bibnamefont{Hawrylak}}, \bibnamefont{and}
  \bibinfo{author}{\bibfnamefont{A.}~\bibnamefont{Wojs}},
  \emph{\bibinfo{title}{Quantum Dots}} (\bibinfo{publisher}{Springer-Verlag},
  \bibinfo{address}{Berlin}, \bibinfo{year}{1998}).

\bibitem[{\citenamefont{Stevenson and
  Kyriakidis}(2011)}]{Stevenson2011Chiral-spin-tex}
\bibinfo{author}{\bibfnamefont{C.~J.} \bibnamefont{Stevenson}}
  \bibnamefont{and}
  \bibinfo{author}{\bibfnamefont{J.}~\bibnamefont{Kyriakidis}},
  \bibinfo{journal}{Phys. Rev. B} \textbf{\bibinfo{volume}{83}},
  \bibinfo{pages}{115306} (\bibinfo{year}{2011}).

\bibitem[{\citenamefont{Dresselhaus}(1955)}]{DRESSELHAUS1955SPIN-ORBIT-COUP}
\bibinfo{author}{\bibfnamefont{G.}~\bibnamefont{Dresselhaus}},
  \bibinfo{journal}{Phys. Rev.} \textbf{\bibinfo{volume}{100}},
  \bibinfo{pages}{580} (\bibinfo{year}{1955}).

\bibitem[{\citenamefont{Bychkov and Rashba}(1984)}]{BYCHKOV1984PROPERTIES-OF-A}
\bibinfo{author}{\bibfnamefont{Y.~A.} \bibnamefont{Bychkov}} \bibnamefont{and}
  \bibinfo{author}{\bibfnamefont{E.~I.} \bibnamefont{Rashba}},
  \bibinfo{journal}{JETP Lett.} \textbf{\bibinfo{volume}{39}},
  \bibinfo{pages}{78} (\bibinfo{year}{1984}).

\bibitem[{\citenamefont{D'yakonov and
  Perel'}(1972)}]{Dyakonov1972Spin-relaxation}
\bibinfo{author}{\bibfnamefont{M.~I.} \bibnamefont{D'yakonov}}
  \bibnamefont{and} \bibinfo{author}{\bibfnamefont{V.~I.}
  \bibnamefont{Perel'}}, \bibinfo{journal}{Sov. Phys. Solid State}
  \textbf{\bibinfo{volume}{13}}, \bibinfo{pages}{3023} (\bibinfo{year}{1972}).

\bibitem[{\citenamefont{de~Andrada~e Silva
  et~al.}(1994)\citenamefont{de~Andrada~e Silva, La~Rocca, and
  Bassani}}]{SILVA1994SPIN-SPLIT-SUBB}
\bibinfo{author}{\bibfnamefont{E.~A.} \bibnamefont{de~Andrada~e Silva}},
  \bibinfo{author}{\bibfnamefont{G.~C.} \bibnamefont{La~Rocca}},
  \bibnamefont{and} \bibinfo{author}{\bibfnamefont{F.}~\bibnamefont{Bassani}},
  \bibinfo{journal}{Phys. Rev. B} \textbf{\bibinfo{volume}{50}},
  \bibinfo{pages}{8523} (\bibinfo{year}{1994}).

\bibitem[{\citenamefont{de~Andrada~e Silva
  et~al.}(1997)\citenamefont{de~Andrada~e Silva, La~Rocca, and
  Bassani}}]{Silva1997Spin-orbit-spli}
\bibinfo{author}{\bibfnamefont{E.~A.} \bibnamefont{de~Andrada~e Silva}},
  \bibinfo{author}{\bibfnamefont{G.~C.} \bibnamefont{La~Rocca}},
  \bibnamefont{and} \bibinfo{author}{\bibfnamefont{F.}~\bibnamefont{Bassani}},
  \bibinfo{journal}{Phys. Rev. B} \textbf{\bibinfo{volume}{55}},
  \bibinfo{pages}{16293} (\bibinfo{year}{1997}).

\bibitem[{\citenamefont{de~Sousa and
  Das~Sarma}(2003)}]{Sousa2003Gate-control-of}
\bibinfo{author}{\bibfnamefont{R.}~\bibnamefont{de~Sousa}} \bibnamefont{and}
  \bibinfo{author}{\bibfnamefont{S.}~\bibnamefont{Das~Sarma}},
  \bibinfo{journal}{Phys. Rev. B} \textbf{\bibinfo{volume}{68}},
  \bibinfo{pages}{155330} (\bibinfo{year}{2003}).

\bibitem[{\citenamefont{Florescu and
  Hawrylak}(2006)}]{Florescu2006Spin-relaxation}
\bibinfo{author}{\bibfnamefont{M.}~\bibnamefont{Florescu}} \bibnamefont{and}
  \bibinfo{author}{\bibfnamefont{P.}~\bibnamefont{Hawrylak}},
  \bibinfo{journal}{Phys. Rev. B} \textbf{\bibinfo{volume}{73}},
  \bibinfo{pages}{045304} (\bibinfo{year}{2006}).

\bibitem[{\citenamefont{Nitta et~al.}(1997)\citenamefont{Nitta, Akazaki,
  Takayanagi, and Enoki}}]{Nitta1997Gate-control-of}
\bibinfo{author}{\bibfnamefont{J.}~\bibnamefont{Nitta}},
  \bibinfo{author}{\bibfnamefont{T.}~\bibnamefont{Akazaki}},
  \bibinfo{author}{\bibfnamefont{H.}~\bibnamefont{Takayanagi}},
  \bibnamefont{and} \bibinfo{author}{\bibfnamefont{T.}~\bibnamefont{Enoki}},
  \bibinfo{journal}{Phys. Rev. Lett.} \textbf{\bibinfo{volume}{78}},
  \bibinfo{pages}{1335} (\bibinfo{year}{1997}).

\bibitem[{\citenamefont{Engels et~al.}(1997)\citenamefont{Engels, Lange,
  Schapers, and Luth}}]{Engels1997Experimental-an}
\bibinfo{author}{\bibfnamefont{G.}~\bibnamefont{Engels}},
  \bibinfo{author}{\bibfnamefont{J.}~\bibnamefont{Lange}},
  \bibinfo{author}{\bibfnamefont{T.}~\bibnamefont{Schapers}}, \bibnamefont{and}
  \bibinfo{author}{\bibfnamefont{H.}~\bibnamefont{Luth}},
  \bibinfo{journal}{Phys. Rev. B} \textbf{\bibinfo{volume}{55}},
  \bibinfo{pages}{R1958} (\bibinfo{year}{1997}).

\bibitem[{\citenamefont{Studer et~al.}(2009)\citenamefont{Studer, Salis,
  Ensslin, Driscoll, and Gossard}}]{Studer2009Gate-Controlled}
\bibinfo{author}{\bibfnamefont{M.}~\bibnamefont{Studer}},
  \bibinfo{author}{\bibfnamefont{G.}~\bibnamefont{Salis}},
  \bibinfo{author}{\bibfnamefont{K.}~\bibnamefont{Ensslin}},
  \bibinfo{author}{\bibfnamefont{D.~C.} \bibnamefont{Driscoll}},
  \bibnamefont{and} \bibinfo{author}{\bibfnamefont{A.~C.}
  \bibnamefont{Gossard}}, \bibinfo{journal}{Phys. Rev. Lett.}
  \textbf{\bibinfo{volume}{103}}, \bibinfo{pages}{027201}
  (\bibinfo{year}{2009}).

\end{thebibliography}

\end{document}